\documentclass[aps,prc,twocolumn,superscriptaddress,showpacs,
nofootinbib]{revtex4-1}
\usepackage{hyperref}
\usepackage{bm}
\usepackage[all]{hypcap}
\usepackage{graphicx,amsmath,amssymb,bm,tabularx}
\usepackage[usenames,dvipsnames,svgnames,table]{xcolor}
\usepackage{enumerate}
\usepackage{isotope}
\usepackage{multirow,dcolumn}
\usepackage{bbm}
\usepackage{braket}
\usepackage{textcomp}
\usepackage{gensymb}

\newcolumntype{.}{D{.}{.}{2.3}}
\newcolumntype{,}{D{.}{.}{1.1}}

\hypersetup{colorlinks=true,citecolor=[rgb]{0,0,0.8},
linkcolor=[rgb]{0,0,0.8},urlcolor=[rgb]{0,0,0.8}}

\allowdisplaybreaks[1]

\newcommand{\fmiq}{~\text{fm}^{-3}}

\newcommand{\mev}{~\text{MeV}}

\renewcommand{\vec}[1]{\mathbf{#1}}

\newcommand{\fm}{~\mathrm{fm}}

\newcommand{\beq}{\begin{equation}}
\newcommand{\eeq}{\end{equation}}

\newcommand{\fet}[1]{\mbox{\boldmath $#1$}}
\definecolor{dartmouth}{rgb}{0.05,0.5,0.06}

\usepackage[normalem]{ulem}
\definecolor{joelcolor}{rgb}{0.90,0.40,0.40}

\allowdisplaybreaks

\begin{document}

\title{Analyzing the Fierz rearrangement freedom for local chiral
two-nucleon potentials}
\author{L.\ Huth}
\email[E-mail:~]{lukashuth@theorie.ikp.physik.tu-darmstadt.de}
\affiliation{Institut f\"ur Kernphysik, Technische Universit\"at Darmstadt, 64289 Darmstadt, Germany}
\affiliation{ExtreMe Matter Institute EMMI, GSI Helmholtzzentrum f\"ur
Schwerionenforschung GmbH, 64291 Darmstadt, Germany}
\author{I.\ Tews}
\email[E-mail:~]{itews@uw.edu}
\affiliation{Institute for Nuclear Theory,
University of Washington, Seattle, Washington 98195-1550, USA}
\affiliation{JINA-CEE, Michigan State University, 
East Lansing, Michigan, 48823, USA}
\author{J.\ E.\ Lynn}
\email[E-mail:~]{joel.lynn@gmail.com}
\affiliation{Institut f\"ur Kernphysik, Technische Universit\"at Darmstadt, 64289 Darmstadt, Germany}
\affiliation{ExtreMe Matter Institute EMMI, GSI Helmholtzzentrum f\"ur
Schwerionenforschung GmbH, 64291 Darmstadt, Germany}
\author{A.\ Schwenk}
\email[E-mail:~]{schwenk@physik.tu-darmstadt.de}
\affiliation{Institut f\"ur Kernphysik, Technische Universit\"at Darmstadt, 64289 Darmstadt, Germany}
\affiliation{ExtreMe Matter Institute EMMI, GSI Helmholtzzentrum f\"ur
Schwerionenforschung GmbH, 64291 Darmstadt, Germany}
\affiliation{Max-Planck-Institut f\"ur Kernphysik, Saupfercheckweg 1, 
69117 Heidelberg, Germany}

\begin{abstract}
Chiral effective field theory is a framework to derive systematic nuclear
interactions. 
It is based on the symmetries of quantum chromodynamics and includes 
long-range pion physics explicitly, while shorter-range
physics is expanded in a general operator basis.
The number of low-energy couplings at a particular order in
the expansion can be reduced by exploiting the fact that nucleons are
fermions and therefore obey the Pauli exclusion principle.
The antisymmetry permits the selection of a subset of the allowed
contact operators at a given order.
When local regulators are used for these short-range interactions,
however, this ``Fierz rearrangement freedom'' is violated.
In this paper, we investigate the impact of this violation at leading order 
(LO) in the chiral expansion.
We construct LO and next-to-leading order (NLO) potentials for all
possible LO-operator pairs and study their reproduction of phase shifts,
the \isotope[4]{He} ground-state energy, and the neutron-matter energy at 
different densities.
We demonstrate that the Fierz rearrangement freedom is partially
restored at NLO where subleading contact interactions enter. 
We also discuss implications for local chiral three-nucleon interactions.
\end{abstract}

\maketitle

\section{Introduction}

Chiral effective field theory (EFT)~\cite{Epelbaum:2008ga, Machleidt:2011zz} 
is a powerful framework to derive nuclear interactions. It is connected to the 
symmetries of quantum chromodynamics (QCD) and provides a systematic
expansion for nuclear forces, including two-nucleon ($NN$) and
many-nucleon interactions, which enables calculations with controlled 
theoretical uncertainties. Chiral EFT explicitly includes long-range 
pion-exchange physics and parametrizes shorter-range physics by 
the most general set of contact operators that is permitted by all 
symmetries of QCD. Thus, in chiral EFT, $NN$ interactions are given 
by the sum of pionic and short-range contributions, 
\begin{align}
V_{NN}(Q,\Lambda, C_i^{(\pi)})=V_{\pi}(Q,\Lambda, C_i^{\pi})+
V_{\text{cont}}(Q,\Lambda, C_i)\,,
\end{align}
where $Q$ denotes the momenta involved or the pion mass and 
$\Lambda$ is a regularization or cutoff scale. Furthermore, $V_\pi$ 
stands for the contribution to the interaction from long-range pion 
exchanges, while $V_\text{cont}$ contains all of the contributions from 
short-range contact operators. The long-range pion exchanges depend 
on a set of pion-nucleon couplings, $C_i^{\pi}$, and the short-range 
interactions depend on a set of low-energy couplings (LECs), $C_i$, that 
are typically fit to experimental data. Considerable effort has been 
invested recently into the improvement of chiral interactions; 
see, e.g., Refs.~\cite{Epelbaum:2014efa, Epelbaum:2014sza,
Ekstrom:2015rta,Carlsson:2015vda, Entem:2017gor}.
 
The most general set of $NN$ contact operators at a particular order in 
the chiral expansion is given by all combinations of spin, isospin, and 
momentum operators that are permitted by the symmetries at this order. 
In addition to 
the spin and isospin of the two nucleons, these operators depend on two 
momenta that can be chosen to be the momentum transfer ${\bf q}={\bf p}'
-{\bf p}$ with the incoming and outgoing relative momenta ${\bf p}=({\bf p}_1
-{\bf p}_2)/2$ and ${\bf p}'=({\bf p}'_1-{\bf p}'_2)/2$, respectively, and the 
momentum transfer in the exchange channel ${\bf k}=\frac12 \left({\bf p}'+ 
{\bf p}\right)$. Interactions that depend only on ${\bf q}$ are local, i.e., they 
depend only on the relative distance ${\bf r}$ upon Fourier transformation to 
coordinate space, while ${\bf k}$ dependencies lead to nonlocal interactions.

The number of contact operators at each order in the chiral expansion can 
be reduced (by a factor of two for $NN$ interactions), due to the fact that 
nucleons are fermions and therefore obey 
the Pauli exclusion principle. As a result of the antisymmetry, only a subset 
of the allowed contact operators at a given order is linearly independent.
Generally only the linearly independent contact operators are therefore
included. While several subsets can be chosen at each 
order, they all lead to the same predictions for physical observables.
This property is known as Fierz rearrangement freedom or
Fierz ambiguity, given the analogy to Fierz identities in four-fermion
interactions~\cite{Fierz1937}.

Recently, local chiral interactions have been developed that can be 
used in quantum Monte Carlo (QMC) methods~\cite{Gezerlis:2013ipa, 
Gezerlis:2014zia, Tews:2015ufa, Piarulli:2014bda,Piarulli:2017dwd}. This has led to the 
first investigations of light nuclei and neutron matter using QMC methods in 
combination with chiral EFT interactions~\cite{Lynn:2014zia,Lynn:2015jua,
Klos:2016fdb, Chen:2016bde,Piarulli:2016vel,Gandolfi:2016bth, 
Lynn:2017fxg,Piarulli:2017dwd}.
However, it has been found that local regulator functions introduce
sizable regulator artifacts in the three-nucleon ($3N$)
sector~\cite{Tews:2015ufa,Lynn:2015jua,Dyhdalo:2016ygz}:
First, one finds less repulsion from the $3N$ two-pion-exchange
interaction than for typical nonlocal regulators, and second, there is
a violation of the Fierz rearrangement freedom for the short-range 
$3N$ contact interactions, i.e., calculations depend on the choice of the 
short-range operators; see also Ref.~\cite{Lovato:2011ij}. While it is true 
that all regulator functions introduce regulator artifacts at finite cutoffs, the 
effects of the violation of the Fierz ambiguity have been found to be larger 
than or comparable to other sources of uncertainty, as demonstrated in
Refs.~\cite{Lynn:2015jua,Dyhdalo:2016ygz}. This additional regulator 
artifact is present for local short-range
regulators~\cite{Navratil:2007zn,Tews:2015ufa}, but not for typical
nonlocal short-range regulators used in previously derived
nonlocal $3N$ interactions~\cite{vanKolck:1994yi,Epelbaum:2002vt}.

A similar violation of the Fierz ambiguity due to local regulators also
appears in the $NN$ sector. Based on power 
counting arguments, one would expect these effects to be even larger 
in the $NN$ than in the $3N$ sector. Thus, it is important to study 
how physical observables are affected by the violation of the Fierz
ambiguity in $NN$ interactions.
Moreover, the $NN$ sector provides an ideal testing ground for
understanding this additional regulator artifact. 

In this paper, we explore the violation of the Fierz ambiguity at leading
order (LO) in the $NN$ sector and investigate the effects of this violation 
on the local coordinate-space interactions of
Refs.~\cite{Gezerlis:2013ipa,Gezerlis:2014zia}.
We study phase shifts, the ground-state energy of \isotope[4]{He} using the 
Green's function Monte Carlo (GFMC) method~\cite{Carlson:2014vla}, 
and the energy per particle of neutron matter
using the auxiliary-field diffusion Monte Carlo (AFDMC) 
method~\cite{Carlson:2014vla} for different short-range operator 
combinations. 
After investigating this regulator artifact at LO, we show that at 
next-to-leading order (NLO) the Fierz rearrangement freedom is 
partially restored and the remaining violation is smaller than typical
chiral uncertainty estimates. 
We emphasize that this effect is not present in nonlocal $NN$ 
interactions~\cite{Ekstrom:2015rta,Carlsson:2015vda,
Entem:2017gor, Entem:2003ft,EGMN3LO} or semilocal 
interactions~\cite{Epelbaum:2014efa, Epelbaum:2014sza}.

This paper is structured as follows.
In Sec.~\ref{seq:fierz} we explain in more detail the Fierz rearrangement 
freedom and 
its relation to different regulator choices.
In Sec.~\ref{seq:locpot} we study the consequences of the violation of 
the Fierz ambiguity for local chiral LO interactions and then investigate
``complete'' LO interactions in Sec.~\ref{sec:completeLO}.
In Sec.~\ref{seq:nlo} we study the effects of the inclusion of NLO
corrections. Finally, we summarize in Sec.~\ref{seq:conclusion}.

\section{Fierz rearrangement freedom} \label{seq:fierz}
The $NN$ interaction at LO ($Q^0$) in Weinberg power 
counting has two contributions: The local
one-pion--exchange (OPE) interaction and momentum-independent
contact interactions. The general set of LO contact interactions
consistent with the symmetries is given by the spin-isospin operators 
$\mathbbm{1}, \sigma_{12}= {\fet \sigma}_1\cdot \fet{\sigma}_2, 
\tau_{12}={\fet \tau}_1\cdot \fet{\tau}_2$, and $\sigma_{12}\tau_{12}$.

\begin{equation}
V_ {\text{cont}}^{(0)} = C_{\mathbbm{1}} +C_{\sigma} \sigma_{12} 
+ C_{\tau} \tau_{12} + C_{\sigma \tau} \sigma_{12}\tau_{12}\,.
\label{eq:ch_cont_LO}
\end{equation}
Even though antisymmetry is a basic symmetry only of the many-body states,
in the following we study the antisymmetrized potential to explain the
origin of the Fierz rearrangement freedom.
The interaction after antisymmetrization, $V_{\text{as}}$, is given by
\begin{equation}
V_{\text{as}}({\bf q},{\bf k}) = \frac12 \left(V({\bf q},{\bf k})-\mathcal{A}[V({\bf q},{\bf k})] \right)\,,
\label{eq:antisym_pot}
\end{equation}
with the antisymmetrizer $\mathcal{A}$ defined via
\begin{align}
\mathcal{A}[V({\bf q},{\bf k})]&=\frac14 (1+\sigma_{12})(1+\tau_{12}) \nonumber \\
& \quad \times V\left({\bf q} \rightarrow -2{\bf k}, {\bf k} \rightarrow -\frac12 {\bf q}\right)\,.
\label{eq:antisymmetrizer}
\end{align}
 
Performing the antisymmetrization for the LO contact interaction
explicitly, we find~\cite{Gezerlis:2014zia}
\begin{align}
V^{(0)}_{\text{cont,as}} &= \frac12 \left(1- \frac14 (1+\sigma_{12})(1+\tau_{12})\right) V_ {\text{cont}}^{(0)} \nonumber \\
&= \left(\frac{3}{8}C_{\mathbbm{1}} -\frac{3}{8}C_{\sigma} -\frac{3}{8}C_{\tau} -\frac{9}{8}C_{\sigma \tau} \right) \mathbbm{1} \nonumber \\
& \quad + \left(-\frac{1}{8}C_{\mathbbm{1}} +\frac{5}{8}C_{\sigma} -\frac{3}{8}C_{\tau} +\frac{3}{8}C_{\sigma \tau} \right) \sigma_{12} \nonumber \\
& \quad + \left(-\frac{1}{8}C_{\mathbbm{1}} -\frac{3}{8}C_{\sigma} +\frac{5}{8}C_{\tau} +\frac{3}{8}C_{\sigma \tau} \right) \tau_{12} \nonumber \\
& \quad + \left(-\frac{1}{8}C_{\mathbbm{1}} +\frac{1}{8}C_{\sigma} +\frac{1}{8}C_{\tau} +\frac{3}{8}C_{\sigma \tau} \right) \sigma_{12}\tau_{12} \nonumber \\
&=\tilde C_S + \tilde C_T \sigma_{12} + \left(-\frac{2}{3}\tilde C_S-\tilde C_T \right) \tau_{12} \nonumber \\
&\quad + \left(-\frac13 \tilde C_S\right) \sigma_{12}\tau_{12}\,.
\label{eq:LO_antisymm}
\end{align}

As can be seen, there are only two independent couplings at LO after
antisymmetrization, and these two couplings contribute to the two possible 
$S$-wave scattering channels.
Thus, only two out of the four operator structures are necessary to
describe the contact physics at LO while the remaining operator structures
are recovered after antisymmetrization.
In principle, any linearly independent combination of two out of the
four contact interactions can be chosen, which is used to reduce the
number of LECs when constructing chiral $NN$ interactions. 

Chiral EFT is a low-momentum theory and, thus, when using chiral 
interactions in few- and many-body calculations, it is necessary to apply 
momentum-dependent regulator functions to cut off high-momentum 
modes that would otherwise lead to divergences. In general, the regulator
function can depend on both momentum scales, $f_R({\bf q},{\bf k})$. 
Let us now consider the preceding argument with a regulator for the
short-range potential included. The Fierz ambiguity is preserved if the 
regulator commutes with the antisymmetrizer, i.e., when
\begin{equation}
\label{eq:FierzCond}
f_R({\bf q},{\bf k})=f_R\left(-2{\bf k},-\frac12{\bf q}\right)\,.
\end{equation}
In this case the regulator is just a prefactor in Eqs.~\eqref{eq:antisym_pot} 
and \eqref{eq:LO_antisymm} and does not affect the antisymmetrization 
procedure: There are still only two independent contact operators at LO.
The condition Eq.~\eqref{eq:FierzCond} is fulfilled only when the
regulator is a symmetric and even function of ${\bf q}$ and ${\bf 2k}$.
This is equivalent to regulating the short-range contact interactions with 
symmetric and even functions of ${\bf p}$ and ${\bf p'}$, as has been done 
for previously derived nonlocal~\cite{Ekstrom:2015rta,Carlsson:2015vda,
Entem:2017gor, Entem:2003ft,EGMN3LO} and semilocal 
interactions~\cite{Epelbaum:2014efa, Epelbaum:2014sza}, which use the 
functional form
\begin{equation} \label{eq:nonlocreg}
f_R({\bf p}, {\bf p}')=
\exp\left[-\left(\frac{{\bf p}\vphantom{'}}{\Lambda}\right)^{2n}\right]
\exp\left[-\left(\frac{{\bf p}'}{\Lambda}\right)^{2n}\right]\,,
\end{equation}
where $n$ takes integer values.

In contrast, local regulators $f_R({\bf q})$ violate the Fierz
rearrangement freedom, because they do not commute with the
antisymmetrizer.
Introducing the momentum exchange operator $\mathcal{P}^{\text{m}}$, 
where $\mathcal{P}^\text{m}f({\bf q},{\bf k})=
f(-2{\bf k},-\tfrac{1}{2}{\bf q})$, the antisymmetrized interaction with
local regulators is given by
\begin{widetext}
\begin{equation}
\begin{split}
\label{eq:LO_antisymm_loc}
&V^{(0,\text{loc})}_{\text{cont,as}}=
\frac12\left(1-\frac{\mathcal{P}^{\text{m}}}{4}
(1+\sigma_{12})(1+\tau_{12})\right)V_ {\text{cont}}^{(0)}f_R({\bf q})\\
&=\mathbbm{1}\left(\frac{C_{\mathbbm{1}}}{2}f_R({\bf q})-
\frac18\left(C_{\mathbbm{1}}+3C_{\sigma}+3C_{\tau}+
9C_{\sigma\tau}\right)f_R(2{\bf k})\right)
+\sigma_{12}\left(\frac{C_{\sigma}}{2}f_R({\bf q})-
\frac18\left(C_{\mathbbm{1}}-C_{\sigma}+3C_{\tau}-
3C_{\sigma\tau}\right)f_R(2{\bf k})\right)\\
&+\tau_{12}\left(\frac{C_{\tau}}{2}f_R({\bf q})-
\frac18\left(C_{\mathbbm{1}}+3C_{\sigma}-C_{\tau}-
3C_{\sigma\tau}\right)f_R(2{\bf k})\right)
+\sigma_{12}\tau_{12}\left(\frac{C_{\sigma\tau}}{2}f_R({\bf q})
-\frac18\left(C_{\mathbbm{1}}-C_{\sigma}-C_{\tau}+
C_{\sigma\tau}\right)f_R(2{\bf k})\right)\,,
\end{split}
\end{equation}
\end{widetext}
where $f_R({\bf q})\neq f_R(2{\bf k})$.
All of the above combinations of LECs are linearly independent and,
thus, the Fierz rearrangement freedom is violated. This violation of the 
Fierz ambiguity is a manifestation of the fact that
introducing a regulator function affects terms beyond the order at which
one is working, and should be corrected when subleading contact
operators are included.
Here we illustrate this explicitly by using a Gaussian local regulator,
$f_R({\bf q})=\exp(-(q/\Lambda)^2)$, such that
$\mathcal{P}^\text{m}f_R({\bf q})=f_R(2{\bf k})=\exp(-4(k/\Lambda)^2)$.
Expressing ${\bf k}$ in terms of ${\bf q}$, ${\bf p}$, and
${\bf p}'$, we can write
\begin{equation}
\begin{split}
\mathcal{P}^\text{m}f_R({\bf q})&=\exp\left(-\frac{q^2}{\Lambda^2}\right)
\exp\left(-\frac{4{\bf p\cdot p'}}{\Lambda^2}\right)\\
&=f_R({\bf q})\left(1-\frac{4{\bf p\cdot p'}}{\Lambda^2}+
\mathcal{O}\left((Q/\Lambda)^4\right)\right)\,.
\end{split} 
\label{eq:f2k}
\end{equation}

Inserting Eq.~\eqref{eq:f2k} into Eq.~\eqref{eq:LO_antisymm_loc}, we 
find
\begin{align}
V^{(0,\text{loc})}_{\text{cont,as}} &= \left(\tilde C_S + \tilde C_T \sigma_{12} + \left(-\frac{2}{3}\tilde C_S-\tilde C_T \right) \tau_{12} \right. \\ \nonumber
&\quad + \left. \left(-\frac13 \tilde C_S\right) \sigma_{12} \tau_{12}\right)f_R({\bf q})+ V_{\text{corr}}^f({\bf p \cdot p'})\,,
\end{align}
where $V_\text{corr}^f({\bf p}\cdot{\bf p}')$ captures the higher-order
effects $\sim~{4{\bf p}\cdot{\bf p}'/\Lambda^2}+
{\mathcal{O}((Q/\Lambda)^4)}$ in Eq.~\eqref{eq:f2k}.
Reexpressing ${\bf p}$ and ${\bf p'}$ in terms of ${\bf q}$ and
${\bf k}$, the first correction term in Eq.~\eqref{eq:f2k} can be
rewritten as 
\begin{align}
-\frac{4 {\bf p\cdot p'}}{\Lambda^2}=
-\frac{4}{\Lambda^2}k^2\!+\!\frac{1}{\Lambda^2}q^2\,.
\end{align}
These operators will be introduced at NLO in chiral EFT, and, 
analogously, the higher-order corrections $\mathcal{O}((Q/\Lambda)^4)$
will be included at next-to-next-to-next-to-leading order (N$^3$LO) 
and beyond.

The higher-order terms $V_{\text{corr}}^f({\bf p \cdot p'})$ depend on 
both the explicit form of the chosen regulator and on the order at which 
one is working. Because the correction terms depend on the 
angles between the nucleons, they contribute to higher partial 
waves (nonlocal regulators only depend on the
magnitudes $p$ and $p'$). As a consequence, while, e.g., the LO contact 
interactions only describe the two $S$-wave channels, 
the use of local regulators leads to a mixing of these contributions into higher partial waves.

More generally, every regulator that respects the Fierz rearrangement 
freedom is an even function of ${\bf p}$ and ${\bf p'}$. Then, every 
regulator that mixes LO physics into partial waves with odd $l$, e.g., 
$P$ waves, and as a consequence contains terms of the form 
$({\bf p} \cdot {\bf p'})^{2n+1}$, violates the Fierz rearrangement 
freedom. Thus, it follows that a violation of the Fierz rearrangement 
freedom is equivalent to a mixing of the $S$-wave contact 
interactions into odd-$l$ partial waves. In contrast, mixing of LO 
contact physics into partial waves with the same $S$ and $T$ 
($\Delta l=2n$) is compatible with the Fierz rearrangement freedom. 

Let us consider the violation of the Fierz ambiguity at LO from the
point of view of a partial-wave decomposition.
We again consider a Gaussian regulator in $\vec{q}$.
This regulator can be rewritten as
\begin{align}
\exp{\left(-\frac{\vec{p}^2 + \vec{p}'^2}{\Lambda^2}\right)}\exp{\left(i\frac{(-2 i\vec{p}\cdot\vec{p}')}{\Lambda^2}\right)}\,,\label{eq:GaussDec}
\end{align}
where the first factor is a typical nonlocal Gaussian regulator. The 
second factor, however, depends on the angle between $\vec{p}$
and $\vec{p}'$. 
Expanding this second exponential function in partial waves, we find
\begin{align}
\exp{\left(i\frac{(-2 i\vec{p}\cdot\vec{p}')}{\Lambda^2}\right)} 
&= 4\pi \sum\limits_{lm} i^{l} j_{l}\left(\frac{-2 i p p'}{\Lambda^2}\right)\nonumber\\
&\quad \times Y^{*}_{lm}(\Omega_{p'})Y_{lm}(\Omega_{p})\,,
\end{align}
with $j_l$ a spherical Bessel function, and $Y_{lm}$ a spherical
harmonic.
We can now compare the radial part for this local regulator to a nonlocal 
LO contact interaction with a Gaussian regulator in a partial-wave 
basis. The nonlocal potential leads to
\begin{align}
\braket{plm|\mathrm{V}^\mathrm{nonloc}_{\text{LO}}(\vec{p},\vec{p}')
|p'l'm'} &= 4 \pi C \delta_{l0} \delta_{l'0} \delta_{m0} \delta_{m'0}\nonumber\\
&\quad \times f_R(p,p')\,, \label{eq:non-local_pwd}
\end{align}
while the local version takes the form
\begin{align}
\braket{plm|\mathrm{V}^\mathrm{loc}_{\text{LO}}(\vec{p},\vec{p}')|p'l'm'} 
= 4 \pi C \delta_{ll'} \delta_{mm'} \nonumber\\
\times i^{l}j_{l}\left(\frac{-2 i p p'}{\Lambda^2}\right)f_R(p,p')\,\label{eq:local_pwd}
\end{align}
with $C$ a constant and
$f_R(p,p') = \exp{\left(\frac{-(p^2+p'^2)}{\Lambda^2}\right)}$.
One can easily see that the nonlocal LO interaction in 
Eq.~\eqref{eq:non-local_pwd} only contributes for $l=l'=0$ ($S$~waves). 
The local interaction of Eq.~\eqref{eq:local_pwd}, on the other hand, 
contributes to each partial wave 
with $l = l'$.

The regulator-induced contribution to all partial waves complicates 
fitting procedures and leads to increased theoretical uncertainties. 
The interaction is, however, 
accompanied by a Bessel function in $p$ and $p'$ that, for increasing 
orbital angular momentum $l$, shifts the mixed contributions to larger 
momenta, where they are suppressed by the regulator $f_R(p,p')$ 
itself. 
Though the
preceding argument was for a specific regulator function, we
 expect that the violation of the Fierz rearrangement freedom has the 
largest effect for partial waves with small orbital angular momenta, 
while large-$l$ partial waves are protected by the angular momentum 
barrier.

\begin{figure}[pt!]
\includegraphics[width=0.99\linewidth]{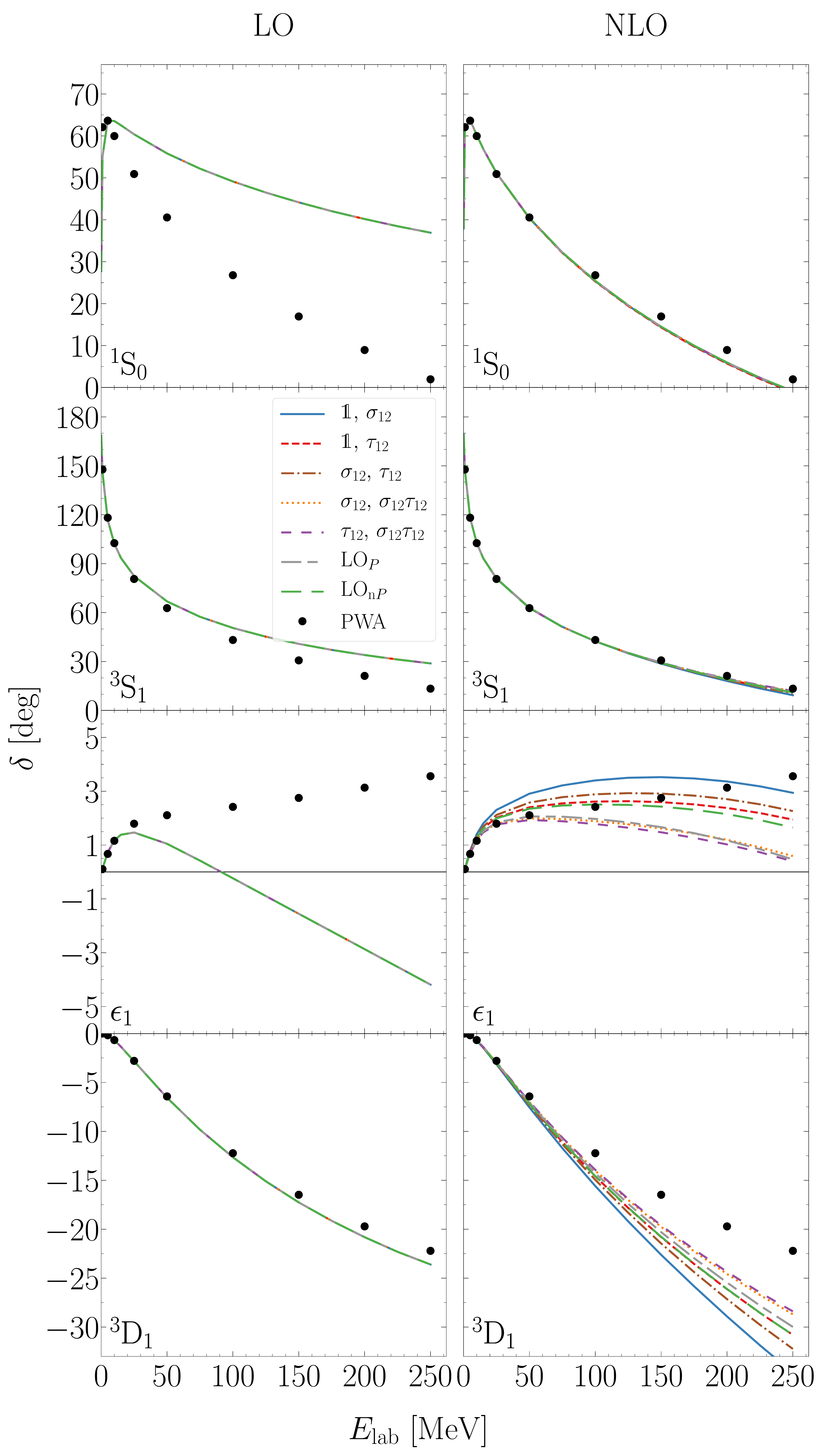}
\caption{\label{fig:SwavePS}
Phase shifts for the $^1S_0$, $^3S_1$, and $^3D_1$ channels and the $J=1$
mixing angle $\epsilon_1$ for $R_0=1.0\fm$ at LO and NLO.
The $S$-wave phase shifts are fit and independent of the choice of the 
LO short-range operators. The lines for different operator choices overlap
except in the case of the mixing angle $\epsilon_1$ and the $^3D_1$
partial wave at NLO.
}
\end{figure} 

\section{Local chiral potentials}\label{seq:locpot}

We now investigate the impact of the violation of Fierz ambiguity on the 
recently developed local chiral $NN$ potentials of Refs.~\cite{Gezerlis:2013ipa,
Gezerlis:2014zia,Piarulli:2014bda,Piarulli:2017dwd}.

In coordinate space, the LO potential is defined by
\begin{align}
V_{NN}^{\text{LO}}(r,R_L,R_S)=V_{\text{OPE}}^{\text{LO}}(r,R_L)+V^{\text{LO}}_{\text{cont}}(r,R_S)\,,
\end{align}
with the long- and short-range coordinate-space cutoffs $R_L$ and $R_S$,
respectively. The OPE interaction has the form
\begin{equation}
\begin{split}
&V_{\text{OPE}}^{\text{LO}}(r,R_L)=\frac{m_\pi^3}{12\pi}
\left(\frac{g_A}{2F_\pi}\right)^2\tau_{12}\, \frac{e^{-m_\pi r}}{m_\pi r}\\
&\quad\times\left[\sigma_{12}+\left(1+\frac{3}{m_\pi r}+
\frac{3}{(m_\pi r)^2}\right)S_{12}\right]f_\text{long}(r,R_L)\,,
\end{split}
\end{equation}
with the pion mass $m_{\pi}=138.03\mev$, the axial coupling constant 
$g_A=1.29$, 
the pion-decay constant $F_{\pi}=92.4\mev$, and the tensor operator 
$S_{12}= 3\fet{\sigma}_1\cdot{\bf \hat{r}}\fet{\sigma}_2\cdot{\bf
\hat{r}}-\sigma_{12}$.
Taking the general set of momentum-independent short-range contact 
operators, we have
\begin{align}
\label{eq:Loc_Cont}
V^{\text{LO}}_{\text{cont}}(r,R_S) &= \left( C_{\mathbbm{1}} + C_{\sigma}\sigma_{12} +C_{\tau}\tau_{12} \right.\\ \nonumber
&\quad \left.+C_{\sigma \tau}\sigma_{12}\tau_{12} \right )
f_\text{short}(r, R_S)\,.
\end{align} 

Below we use the long- and short-range local regulator 
functions
\begin{subequations}
\begin{align}
f_\text{long}(r,R_0)&=1-e^{-(r/R_0)^4}\,,\\
f_\text{short}(r,R_0)&=\frac{e^{-(r/R_0)^4}}{\pi\Gamma(3/4)R_0^3}\,,
\end{align}
\end{subequations}
with a single cutoff scale $R_L=R_S=R_0$ as in
Refs.~\cite{Gezerlis:2013ipa,Gezerlis:2014zia}. Our conclusions apply 
equally well to the long- and short-range local regulator functions of 
Refs.~\cite{Piarulli:2014bda,Piarulli:2017dwd}.

If the Fierz rearrangement freedom is respected, physical observables
are independent of the choice of any linearly independent set of two out of the four
operators in Eq.~\eqref{eq:Loc_Cont}. For local regulators, this is not 
the case and we investigate this effect by constructing LO potentials 
for all possible pairs of contact operators.
In each case, we fit the two LECs to phase shifts in the two $S$-wave 
channels.
More precisely, we fit the spin-isospin LECs to these phase shifts, which we
label according to the spin and isospin quantum numbers, $C_{ST}$,
instead of using the standard partial-wave notation, $C_{^{2S+1}L_J}$.
We reconstruct the operator LECs of Eq.~\eqref{eq:Loc_Cont} according to
\begin{align}
\begin{pmatrix} C_{00}\\ C_{01} \\ C_{10} \\ C_{11} \end{pmatrix} 
= \begin{pmatrix} 1 & -3 & -3 & \hphantom{-}9 \\ 1 & -3 & \hphantom{-}1 & -3 \\ 1 & \hphantom{-}1 & -3 & -3 \\ 1 & \hphantom{-}1 & \hphantom{-}1 & \hphantom{-}1 \end{pmatrix}
\begin{pmatrix} C_{\mathbbm{1}\hphantom{\sigma}}\\ C_{\sigma\hphantom{\sigma}} \\ C_{\tau\hphantom{\sigma}} \\ C_{\sigma \tau} \end{pmatrix}\,.
\label{eq:LECtrafo}
\end{align}
Note that we exclude the pair $\mathbbm{1},\sigma_{12}\tau_{12}$ 
from consideration, because the operators of this pair are linearly dependent
in the two $S$-wave channels.

We fit the LECs $C_{01}$ and $C_{10}$ to the $^1S_0$ and $^3S_1$ 
neutron-proton phase shifts from the partial-wave analysis of
Ref.~\cite{Stoks:1993tb} (PWA) for cutoffs in the range $R_0=1.0-1.2\fm$. 
While fitting to phase shifts contains inherent drawbacks, our goal in 
this work is not to produce high-precision potentials. Nevertheless,
we plan to fit to scattering data in future work.
We use an existing automatic differentiation package~\cite{Lee:AD} 
for Python to obtain numerical gradients for the fits, and feed these into 
Python's BFGS minimization routine. This algorithm is a quasi-Newton 
method, named after its founders Broyden, Fletcher, Goldfarb, and 
Shanno~\cite{BFGS}. We perform least-square optimizations 
to minimize a $\chi^2$ value with respect to the LECs. The $\chi^2$ 
value is defined by
\begin{align} 
\chi^2 = \sum\limits_i^\text{data set} \left(\frac{\delta^\mathrm{PWA}_i - \delta^\mathrm{theo}_i}{\Delta \delta_i}\right)^2\,,
\end{align}
where the uncertainty $\Delta \delta_i$ is given by
\begin{equation}
\Delta \delta_i^2=(\Delta \delta_i^{\text{PWA}})^2+(\Delta \delta_i^{\text{theo}})^2+(\Delta \delta_i^{\text{num}})^2\,.
\end{equation}
Here $\Delta \delta_i^{\text{PWA}}$ is the uncertainty from the 
PWA, $\Delta \delta_i^{\text{theo}}$ is the theoretical model uncertainty 
for the chiral interactions, and $\Delta \delta_i^{\text{num}}$ is due to 
numerical errors. For the theoretical uncertainty we use the relative 
uncertainty and multiply it with a constant $C$ to obtain a
dimensionless $\chi^2$. This is similar to the proposed uncertainty of 
Ref.~\cite{Carlsson:2015vda}. We use $Q =\mathrm{max} 
\left(m_\pi,\,P \right)$, where $P$ is a typical momentum scale of the 
system, and obtain for the LO and NLO uncertainties
\begin{align}
\Delta \delta_i^{\mathrm{theo, LO}} &= \left(\frac{Q}{\Lambda}\right)^2 C\,, \\
\Delta \delta_i^{\mathrm{theo, NLO}} &= \left(\frac{Q}{\Lambda}\right)^3 C\,,
\end{align}
where we take $\Lambda=500$~MeV (400 MeV) for $R_0 = 1.0\fm$ 
(1.2 fm) (see Ref.~\cite{Lynn:2017fxg}) and $C=1\degree$. When studying 
\isotope[4]{He}, we choose the momentum scale associated with the 
average density in \isotope[4]{He}, $P~\approx~290 \mev$, and when 
studying neutron matter,
we choose $P$ to be the average momentum in a Fermi gas, 
$P= \sqrt{3/5} k_F$, with the Fermi momentum $k_F$. The PWA and 
numerical uncertainties in $\Delta \delta_i$ are negligible compared to this 
theoretical model uncertainty.

\begin{table*}[t!]
\caption{Operator LECs and spin-isospin LECs $C_{00}$ and $C_{11}$ for 
all investigated LO operator combinations. The cutoff $R_0$ is given
in fm and the LEC values are given in fm$^2$. The spin-isospin LECs
$C_{01}\, (C_{10}$) are 
$-1.831\fm^2 \, (-0.317\fm^2)$ for $R_0=1.0\fm$
and
$-2.216\fm^2\, (-1.579\fm^2)$ for $R_0=1.2\fm$.
}
{\renewcommand{\arraystretch}{1.3}
\begin{ruledtabular}
\begin{tabular}{c,......}
\label{tab:LOLECs}
Operators&\multicolumn{1}{c}{$R_0$}&
\multicolumn{1}{c}{$C_{\mathbbm{1}}$}&\multicolumn{1}{c}{$C_{\sigma}$}&
\multicolumn{1}{c}{$C_{\tau}$}&\multicolumn{1}{c}{$C_{\sigma\tau}$}&
\multicolumn{1}{c}{$C_{00}$}&\multicolumn{1}{c}{$C_{11}$}\\
\hline \\
\multirow{2}{*}{$\mathbbm{1},\,\sigma_{12}$}& 1.0 &-0.696& 0.378& 0.000& 0.000& -1.831& -0.317\\
& 1.2 &-1.738& 0.159& 0.000& 0.000& -2.216& -1.579\\
\multirow{2}{*}{$\mathbbm{1},\,\tau_{12}$}& 1.0 &-1.452& 0.000& -0.378& 0.000& -0.317& -1.831\\
& 1.2 &-2.057& 0.000& -0.159& 0.000& -1.579& -2.216\\
\multirow{2}{*}{$\sigma_{12},\,\tau_{12}$}& 1.0 &0.000& 0.726& 0.348& 0.000& -3.222& 1.074\\
& 1.2 &0.000& 1.028& 0.869& 0.000& -5.693& 1.898\\
\multirow{2}{*}{$\sigma_{12},\,\sigma_{12}\tau_{12}$}& 1.0 &0.000& 0.378& 0.000& 0.232& 0.951& 0.610\\
& 1.2 &0.000& 0.159& 0.000& 0.579& 4.737& 0.739\\
\multirow{2}{*}{$\tau_{12},\,\sigma_{12}\tau_{12}$}& 1.0 &0.000& 0.000& -0.378& 0.484& 5.492& 0.106\\
& 1.2 &0.000& 0.000& -0.159& 0.686& 6.649& 0.526\\
\multirow{2}{*}{$\text{LO}_P$}& 1.0 &-1.980& -0.415& -0.793& 0.101& 2.548& -3.087\\
& 1.2 &-2.208 & -0.346 & -0.505 & 0.180 & 1.962 & -2.879\\
\multirow{2}{*}{$\text{LO}_{\text{n}P}$}& 1.0 &-0.403& 0.323& -0.055& 0.134& 0.000& 0.000\\
& 1.2 &-0.712& 0.317& 0.158& 0.237& 0.000& 0.000
\end{tabular}
\end{ruledtabular}}
\end{table*}

\begin{table*}[t!]
\caption{Operator LECs for 
all investigated NLO operator combinations. The cutoff $R_0$ is given
in$\fm$ and the LEC values are given in fm$^2$ for $C_{\mathbbm{1}}$, $C_{\sigma}$, $C_{\tau}$, and, $C_{\sigma\tau}$ and in fm$^4$ for $C_1$ -- $C_7$. 
}
{\renewcommand{\arraystretch}{1.0}
\begin{ruledtabular}
\begin{tabular}{c,...........}
\label{tab:NLOLECs}
Operators&\multicolumn{1}{c}{$R_0$}&
\multicolumn{1}{c}{$C_{\mathbbm{1}}$}&\multicolumn{1}{c}{$C_{\sigma}$}&
\multicolumn{1}{c}{$C_{\tau}$}&\multicolumn{1}{c}{$C_{\sigma\tau}$}&
\multicolumn{1}{c}{$C_{1}$}&\multicolumn{1}{c}{$C_{2}$}&
\multicolumn{1}{c}{$C_{3}$}&\multicolumn{1}{c}{$C_{4}$}&
\multicolumn{1}{c}{$C_{5}$}&\multicolumn{1}{c}{$C_{6}$}&
\multicolumn{1}{c}{$C_{7}$}\\
\hline \\
\multirow{2}{*}{$\mathbbm{1},\,\sigma_{12}$}& 1.0 &2.080& 1.036& 0.000& 0.000& 0.285& 0.184& -0.120& 0.063& -2.241& 0.304& -0.286\\
& 1.2 &0.080& 0.729& 0.000& 0.000& 0.212& 0.209& -0.156& 0.092& -2.112& 0.343& -0.376\\
\multirow{2}{*}{$\mathbbm{1},\,\tau_{12}$}& 1.0 &-0.830& 0.000& -0.192& 0.000& 0.102& -0.000& -0.149& -0.049& -1.373& 0.171& -0.102\\
& 1.2 &-1.489& 0.000& -0.613& 0.000& -0.104& 0.051& -0.298& 0.035& -1.285& 0.391& -0.293\\
\multirow{2}{*}{$\sigma_{12},\,\tau_{12}$}& 1.0 &0.000& 0.296& -0.144& 0.000& 0.174& 0.061& -0.131& -0.012& -1.698& 0.205& -0.161\\
& 1.2 &0.000& 0.518& -0.562& 0.000& 0.076& 0.216& -0.255& 0.139& -1.923& 0.443& -0.461\\
\multirow{2}{*}{$\sigma_{12},\,\sigma_{12}\tau_{12}$}& 1.0 &0.000& -0.133& 0.000& 0.477& 0.280& -0.027& -0.109& -0.026& -1.743& 0.093& -0.052\\
& 1.2 &0.000& 0.054& 0.000& 0.651& 0.381& 0.028& -0.131& 0.058& -2.079& 0.149& -0.179\\
\multirow{2}{*}{$\tau_{12},\,\sigma_{12}\tau_{12}$}& 1.0 &0.000& 0.000& 0.173& 0.402& 0.292& -0.016& -0.091& -0.036& -1.792& 0.083& -0.045\\
& 1.2 &0.000& 0.000& -0.071& 0.681& 0.373& 0.023& -0.142& 0.064& -2.059& 0.155& -0.183\\
\multirow{2}{*}{$\text{LO}_P$}& 1.0 &3.517& 0.409& 0.414& 1.246& 0.562& 0.053& -0.059& 0.054& -2.524& 0.113& -0.101\\
& 1.2 &0.843& -0.236& -0.364& 1.101& 0.473& 0.024& -0.159& 0.118& -2.201& 0.168& -0.206\\
\multirow{2}{*}{$\text{LO}_{\text{n}P}$}& 1.0 &-0.254& 0.173& -0.004& 0.085& 0.191& 0.027& -0.116& -0.032& -1.667& 0.156& -0.111\\
& 1.2 &-0.717& 0.290& 0.189& 0.239& 0.234& 0.067& -0.109& -0.004& -1.927& 0.170& -0.187\\
\end{tabular}
\end{ruledtabular}}
\end{table*}

\begin{figure}[t]
\centering
\includegraphics[width=\columnwidth]{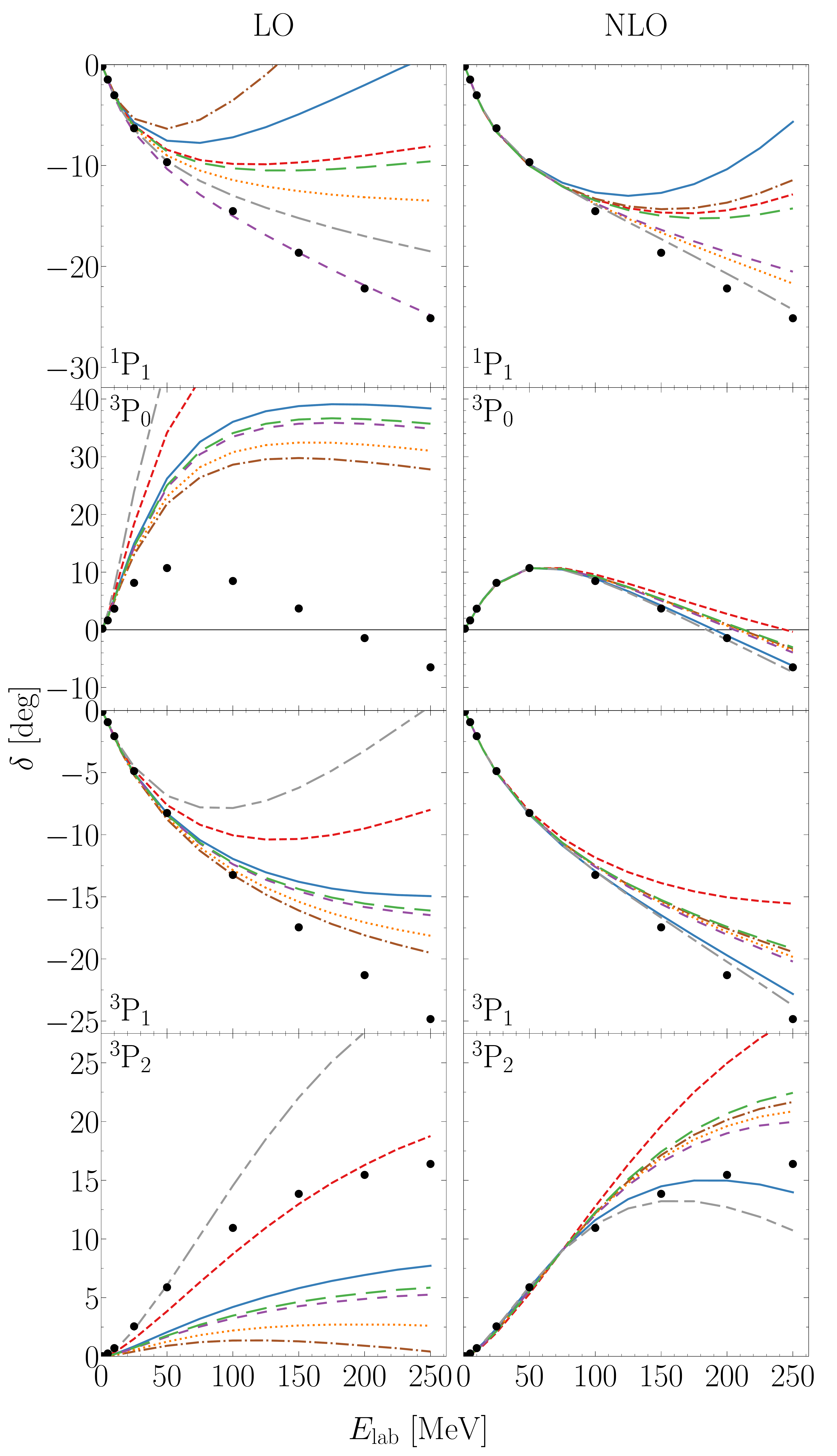}
\caption{\label{fig:PwavePS}
Phase shifts for the $^1P_1$ and triplet $^3P_J$ waves for 
$R_0=1.0\fm$. The legend is as in Fig.~\ref{fig:SwavePS}. }
\end{figure}

\begin{figure}[t]
\centering
\includegraphics[width=0.99\linewidth]{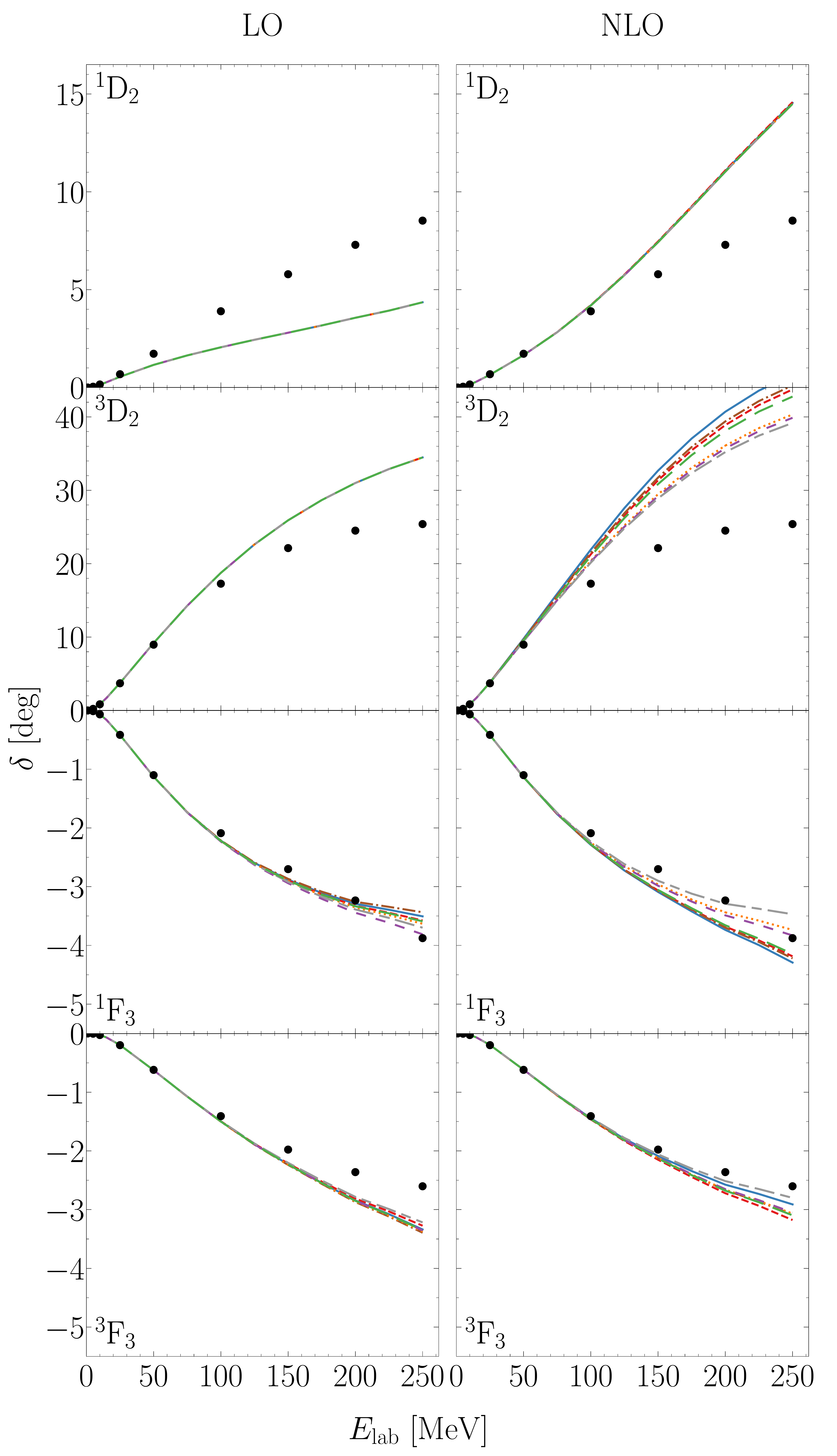}
\caption{\label{fig:DwavePS}
Phase shifts for the $^1D_2$, $^3D_2$, $^1F_3$, and $^3F_3$ 
partial waves for $R_0=1.0\fm$. The legend is as in
Fig.~\ref{fig:SwavePS}. }
\end{figure} 

We fit the interactions up to energies of $50\mev$ at LO and up to 
$150\mev$ at NLO. In particular, we fit to the energy points:
\pagebreak
\begin{align*}
\mathrm{LO\!:}\,\, &1,\,5,\,10,\,25,\,50\text{ MeV}\,,\\
\mathrm{NLO\!:}\,\, &1,\,5,\,10,\,25,\,50,\,100,\,150\text{ MeV}\,.
\end{align*}
For all operator pairs, we present the fitted operator LECs, as well 
as the spin-isospin LECs $C_{00}$ and $C_{11}$ in Table~\ref{tab:LOLECs}
and show the phase shifts at LO in the left panels of
Figs.~\ref{fig:SwavePS}--\ref{fig:DwavePS}. The NLO results shown in 
the right panels will be discussed in Sec.~\ref{seq:nlo}. 

\begin{figure}[t]
\centering
\includegraphics[width=\columnwidth]{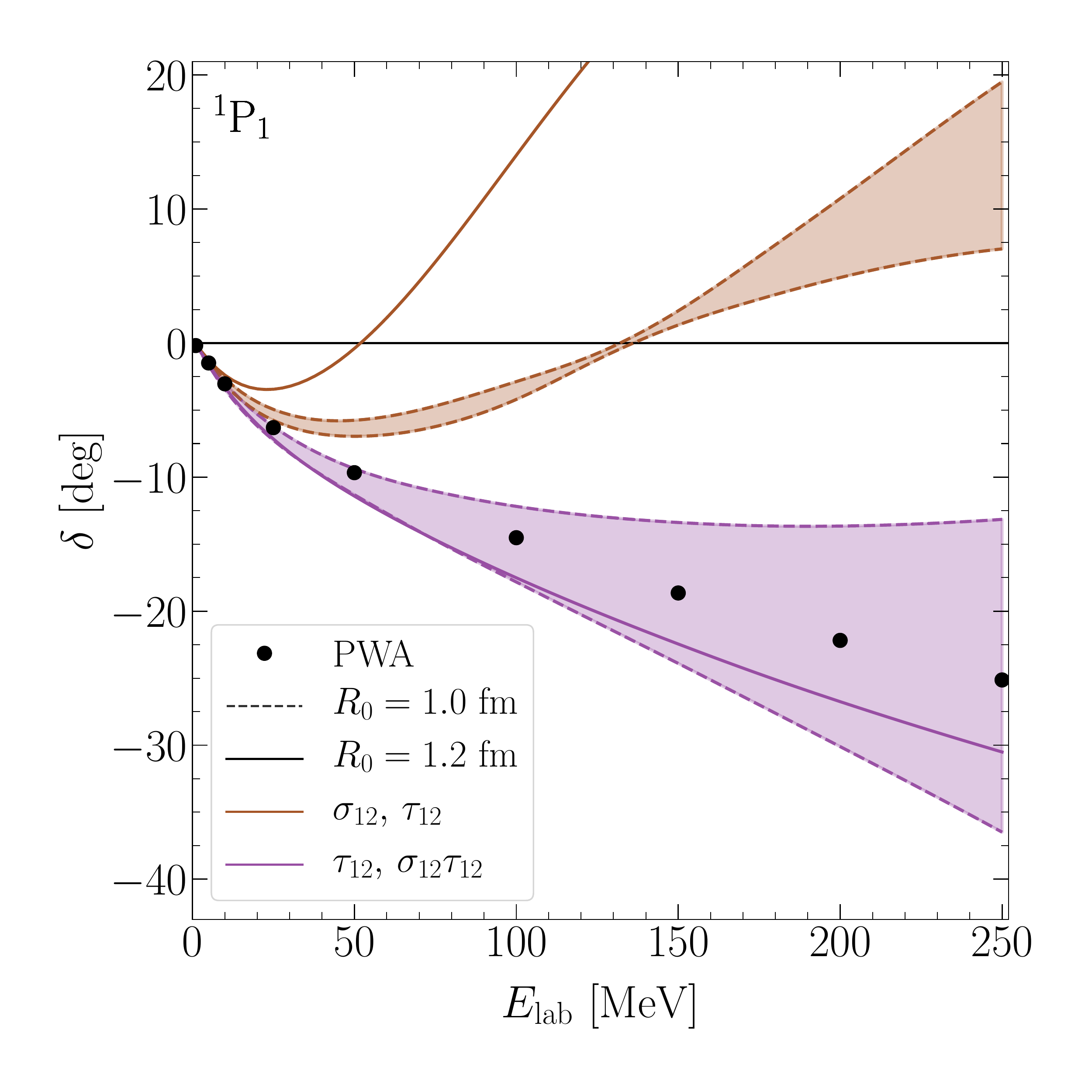}
\caption{\label{fig:1P1}
Phase shift as a function of laboratory energy in the $^1P_1$ channel 
for the two interactions that give the 
most extreme results: the $\sigma_{12},\tau_{12}$ and 
$\tau_{12},\sigma_{12} \tau_{12}$ interactions. We show the results
for $R_0=1.0 \fm$ with EKM uncertainty estimates (bands, see text) 
and the results for $R_0=1.2 \fm$ as solid lines. The color scheme is 
as in Fig.~\ref{fig:SwavePS}.
}
\end{figure}

We begin by examining the $^1S_0$, $^3S_1$, and $^3D_1$ phase 
shifts as well as the $J=1$ mixing angle in Fig.~\ref{fig:SwavePS} for 
$R_0=1.0\fm$, in comparison to the PWA. The results for $R_0=1.2\fm$ 
are qualitatively similar; see Fig.~\ref{fig:1P1} for an example
for the $^1P_1$ wave.
The spin-isospin LECs $C_{01}$ and $C_{10}$ are fit and, thus, 
independent of the operator choice. Because these LECs enter the two
$S$-wave and the $^3D_1$ phase shifts, we do not observe any 
dependence on the operator choice. The lines for all operator pairs 
overlap for the $^1S_0$, $^3S_1$, and $^3D_1$ channels and the 
mixing angle. The obtained phase shifts are very close to the corresponding 
phase shifts of Ref.~\cite{Gezerlis:2014zia}.
Furthermore, for all potentials we find the same deuteron binding energy of
$E_d=1.942\mev \, ( 1.949\mev)$ for $R_0=1.0\fm$ ($1.2 \fm$).

While $C_{01}$ and $C_{10}$ do not depend on the chosen operator 
structure, it is immediately clear from Eq.~\eqref{eq:LECtrafo} that 
the operator LECs as well as the spin-isospin LECs $C_{00}$ and $C_{11}$
do.
When comparing the $C_{00}$ and $C_{11}$ values for two different 
operator combinations in Table~\ref{tab:LOLECs}, one can 
see that the Fierz rearrangement freedom is violated for local regulators.
Considering, e.g., the operator combination $\mathbbm{1},\sigma_{12}$
one obtains $C_{00}=C_{01}$ and $C_{11}=C_{10}$.
For the combination $\mathbbm{1},\tau_{12}$, instead, one finds
$C_{00}=C_{10}$ and $C_{11}=C_{01}$. 
As a consequence, the phase shifts in the corresponding spin-isospin
channels are not independent of the operator choice when local
regulators are used.

We show the phase shifts in the $^1P_1$ (determined by $C_{00}$) 
and in the $^3P_J$ (determined by $C_{11}$) partial waves at LO in 
the upper panels of Fig.~\ref{fig:PwavePS} for $R_0=1.0\fm$.
At low energies, the results agree very well with the 
PWA, and different choices for the operator pair lead to the same
phase shifts, as expected.
At higher energies, however, the subleading corrections
become more important and the phase shifts begin to disagree
considerably with the PWA as well as with each 
other.
For all $P$ waves, at energies above $\approx 20\mev$,
the choice of the operator pair clearly affects the phase-shift prediction 
and we observe a large variation for the resulting phase shifts. This
variation follows the ordering expected from Table~\ref{tab:LOLECs}. 
In the $^1P_1$ partial wave, e.g., the variation ranges from 
results for the $\tau_{12},\sigma_{12}\tau_{12}$ interaction,
which describe the PWA results very well, to results  for the 
$\sigma_{12},\tau_{12}$ interaction that even change sign at 
$130\mev$.

Because the $^1P_1$ phase shift experiences a sizable effect, 
we show it again in Fig.~\ref{fig:1P1} for the two operator pairs 
that give the extreme results and for both cutoffs. For $R_0=1.0 \fm$, 
we additionally show the uncertainties according to the prescription 
of Ref.~\cite{Epelbaum:2014efa} (EKM uncertainties) at LO. We observe 
that the violation of the Fierz ambiguity is slightly worse for $R_0=1.2\fm$,
which is expected based on the corresponding smaller momentum-space 
cutoff and,
thus, larger correction terms. We also find, that the uncertainty due to 
the violation of the Fierz ambiguity is probed neither by varying the
cutoff, which has been regarded as a tool for assessing the uncertainty
due to neglecting higher-order contact operators, nor by the EKM
uncertainties. 

In Fig.~\ref{fig:DwavePS} we show phase shifts for the $^1D_2$, $^3D_2$, 
$^1F_3$, and $^3F_3$ partial waves for
$R_0~=~1.0\fm$. For the $D$ waves, only the $S$-wave spin-isospin LECs 
enter, and, thus, the LO phase shifts are independent of the short-range 
operators. For the $F$-wave phase shifts, the dependence on the 
short-range operator structure is nonnegligible only at energies larger 
than $200\mev$, because the higher-$l$ phase shifts are protected by the 
angular-momentum barrier. These findings are consistent with our 
expectations based on the discussion around 
Eqs.~\eqref{eq:non-local_pwd} and \eqref{eq:local_pwd}.

\begin{figure*}[t]
\centering
\includegraphics[width=2.05\columnwidth]{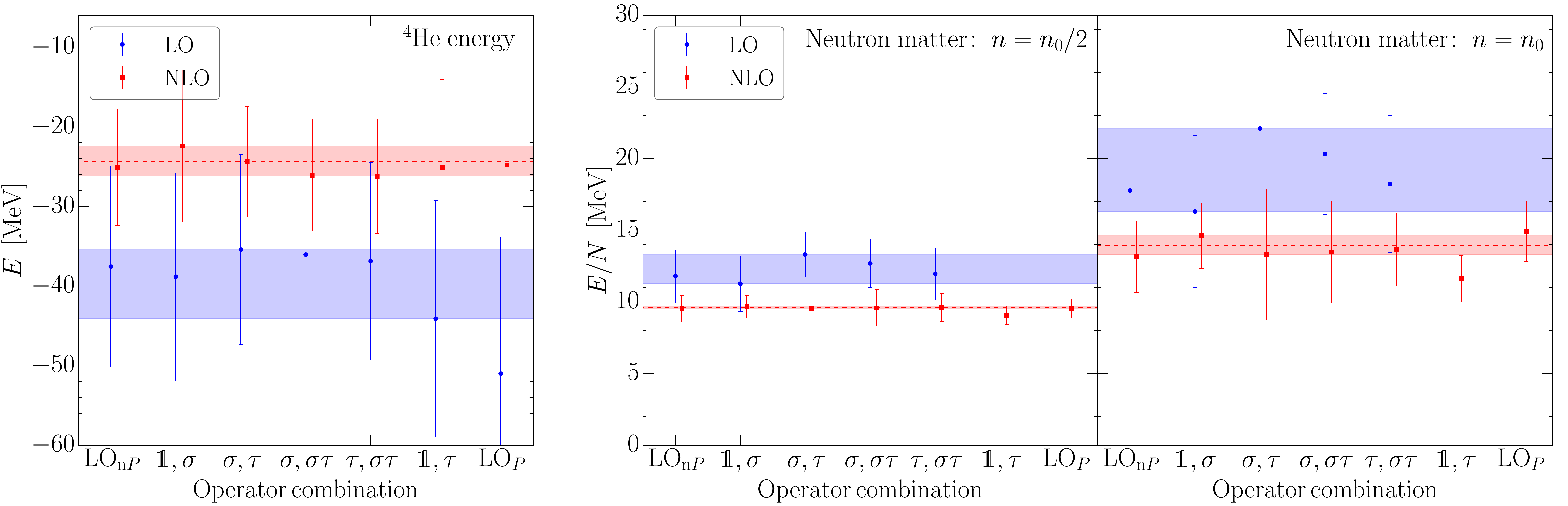}
\caption{\label{fig:NM4He}
Many-body observables at LO and NLO for different LO operator
combinations.
Here we drop the operator subscripts for legibility; i.e.
$\sigma_{12}\to\sigma$, etc.
The shaded regions show the spread for the various operator choices,
while the dashed line denotes the centroid.
When calculating the band and the centroid, we exclude the complete
potentials in all cases and the $\mathbbm{1},\tau$ operator combination
in the case of neutron matter.
Note that for neutron matter at LO, we only show the results for five
interactions, because we find bound neutron matter for the $\text{LO}_P$
and the $\mathbbm{1},\tau_{12}$ interactions. 
Left panel: Ground-state energy of \isotope[4]{He}.
Middle panel: Neutron matter energy per particle at one half saturation density.
Right panel: Neutron matter energy per particle at saturation density.}
\end{figure*}

In addition to $NN$ scattering phase shifts, we investigate the impact 
of different operator choices on other physical observables. In particular,
we study the \isotope[4]{He} ground-state energy $E$ using the GFMC 
method, and the neutron-matter energy per particle $E/N$ at different 
densities using the AFDMC method.
The results are shown in Fig.~\ref{fig:NM4He} for all operator combinations, 
together with the corresponding EKM uncertainties as error bars. We also 
show the spread of all operator pairs as shaded regions. If the Fierz 
rearrangement freedom were respected, the spread would vanish. 

For the \isotope[4]{He} ground-state energy, we observe a sizable dependence 
on the LO operator choice. The \isotope[4]{He} ground-state energy varies
between $-35.4\mev$ and $-44.1\mev$, the $\mathbbm{1},\tau_{12}$ 
interaction being an outlier due to strong $P$-wave attraction (excluding 
this operator choice, the ground-state energy only varies between $-35.4\mev$ 
and $-38.8\mev$.) The spread for the different operator combinations is 
$\sim9\mev$ and comparable to the EKM uncertainties at this order.

In the case of neutron matter, the energy per neutron at half saturation
density $n_0/2$ $(\text{at }n_0)$, with $n_0=0.16 \fmiq$,
ranges between $11.2$ $(16.3)\mev$ for the pair $\mathbbm{1}, 
\sigma_{12}$ and $13.3$ $(22.1)\mev$ for the pair $\sigma_{12},
\tau_{12}$. The spread due to the violation of Fierz rearrangement 
freedom is again comparable to the size of the EKM uncertainties for these 
interactions, which has also been observed for the leading $3N$ contact 
interactions~\cite{Lynn:2015jua}. Furthermore, it is important to note that 
the operator choice $\mathbbm{1}, \tau_{12}$ leads to bound neutron matter
for both densities. This can be easily understood from Table~\ref{tab:LOLECs}:
the spin-isospin LEC entering the triplet $P$ waves is of the same size as 
the $S$-wave LEC and attractive for these operators. We exclude this 
interaction when showing the horizontal band.

\section{``Complete'' LO potentials}\label{sec:completeLO}
 
As we have seen in the previous section, the choice of the LO 
operators clearly impacts the results for phase shifts and the
energies of nuclei and neutron matter, and leads to a range of
results that is not covered by typical uncertainty estimates. To 
correct for this regulator artifact already at LO one could also explicitly 
compute the correction terms and include these in the calculation.
Because it is nontrivial to include nonlocal terms in QMC 
simulations, we will not pursue this approach here.

Instead, we follow an idea similar to the one used in
Ref.~\cite{Lynn:2015jua} and 
construct a LO potential with a projector on $S\neq T$ partial waves.
To implement this projector, we construct a complete LO potential,
i.e., a potential that includes all four LO contact operators. In addition 
to fitting the $^1S_0$ and $^3S_1$ phase shifts (fitting 
$C_{01}$ and $C_{10}$), we enforce that the contribution 
of $V^{\text{LO}}_{\text{cont}}(r,R_0)$ vanishes in the partial waves 
with $T=S$, i.e.,
\begin{align}
\label{eq:Pwave0}
C_{00}&=C_{\mathbbm{1}} -3 C_{\sigma} -3 C_{\tau} +9C_{\sigma \tau}=0\, , \\ \nonumber
C_{11}&=C_{\mathbbm{1}} + C_{\sigma} + C_{\tau} +C_{\sigma \tau}=0\, ,
\end{align}
where we have used Eq.~\eqref{eq:LECtrafo}.
Enforcing the two conditions in Eq.~\eqref{eq:Pwave0} eliminates 
the mixing into $P$ waves. In the following, we call this interaction 
$\text{LO}_{\text{n}P}$, for ``no $P$-wave'' mixing. This potential
eliminates the regulator artifacts in odd-$l$ partial waves and, hence,
is closest to an LO potential that respects the Fierz rearrangement 
freedom. Furthermore, it simplifies the fitting procedure for local
chiral interactions at higher orders in the chiral expansion. This 
potential leads to a good reproduction of the $P$-wave phase shifts, 
and, for $R_0=1.0 \fm$, gives a \isotope[4]{He} ground-state 
energy of $-37.6\mev$, and a neutron-matter energy of $11.8$ 
$(17.8)\mev$ at $n_0/2$ $(n_0)$. Results for this potential are also 
shown in Figs.~\ref{fig:SwavePS}--\ref{fig:DwavePS} and \ref{fig:NM4He}.

In addition to the $\text{LO}_{\text{n}P}$ potential, we also investigated 
a second complete LO potential, where we fit the additional couplings 
to the $^1P_1$ and $^3P_2$ partial waves. We call this interaction 
$\text{LO}_P$. In contrast to the $\text{LO}_{\text{n}P}$ potential, 
the $\text{LO}_P$ potential does not eliminate any regulator artifacts 
at LO but instead matches them to reproduce two $P$-wave phase shifts. 
This potential, however, is too attractive in the triplet $P$ waves, see 
Table~\ref{tab:LOLECs}, and performs worst in the $^3P_0$ and $^3P_1$ 
partial waves. We also investigated the alternatives of fitting this
potential to the $^1P_1$ and one of the other $^3P_J$ partial waves.
These lead to an excellent description of the $^3P_J$ wave under
consideration, but an even worse reproduction of the 
other two triplet $P$ waves, with, for example,
$C_{11}\approx5C_{01}$ in the fit to the $^1P_1$ and $^3P_2$ partial
waves. 

Due to the strong attraction in the triplet $P$ waves, we find a large 
\isotope[4]{He} ground-state energy of $\sim-51\mev$ for the
$\text{LO}_P$ interaction.
Furthermore, this potential leads to bound neutron matter.
The $\text{LO}_{P}$ potential, thus, performs worst in all systems that
we studied.
Again, results for this potential are shown in
Figs.~\ref{fig:SwavePS}--\ref{fig:DwavePS} and \ref{fig:NM4He}.

\section{Next-to-leading order} \label{seq:nlo}

As stated above, the violation of the Fierz rearrangement freedom 
due to the local regulator will be corrected by higher-order terms. 
The first correction is of the order $(Q/\Lambda)^2$ and appears 
at NLO in chiral EFT. In the following, we 
investigate to what extent the subleading NLO short-range operators 
restore the Fierz rearrangement freedom at LO.
 
At NLO, the interactions include momentum-dependent short-range
contact interactions and two-pion--exchange interactions. For 
the contact interactions, the most general set of operators in momentum
space is given by 14 different terms,
\begin{equation}
\begin{split}
V_{\rm cont}^{(2)}&=\gamma_1q^2+\gamma_2q^2\sigma_{12}+\gamma_3q^2\tau_{12}+\gamma_4q^2\sigma_{12}\tau_{12}\\
&+\gamma_5k^2+\gamma_6k^2\sigma_{12}+\gamma_7k^2\tau_{12}+\gamma_8k^2\sigma_{12}\tau_{12}\\
&+\gamma_9({\bm \sigma}_1+{\bm \sigma}_2)\cdot({\bf q}\times{\bf k})+\gamma_{10}({\bm \sigma}_1+{\bm \sigma}_2)\cdot({\bf q}\times{\bf k})\tau_{12}\\
&+\gamma_{11}({\bm \sigma}_1\cdot {\bf q})({\bm \sigma}_2\cdot{\bf q})+\gamma_{12} ({\bm \sigma}_1\cdot {\bf q})({\bm \sigma}_2\cdot{\bf q})\tau_{12}\\
&+\gamma_{13}({\bm \sigma}_1\cdot {\bf k})({\bm \sigma}_2\cdot{\bf k})+\gamma_{14} ({\bm \sigma}_1\cdot {\bf k})({\bm \sigma}_2\cdot{\bf k})\tau_{12}\,.
\label{eq:NLOlong}
\end{split}
\end{equation}
As before, only seven out of these 14 operators are independent for reasons
of antisymmetry if no regulator is included.
We use the same local regulators at NLO as we did at LO.
Then, the Fierz rearrangement freedom is again violated at NLO. In the following, however, 
we neglect the new regulator artifacts that originate at NLO and instead 
focus only on the LO regulator artifacts because they are the largest in size: 
The corrections to the NLO violation appear only at N$^3$LO. Furthermore, 
the NLO regulator artifacts in $S$ and $P$ waves can be absorbed into the LECs and the first
artifacts then appear in $D$ waves, where they are additionally
suppressed by the angular momentum barrier.

In the following, we construct NLO interactions for all possible
pairs of LO operators, as well as the two potentials with the complete
set of LO operators ($\text{LO}_{\text{n}P}$ and $\text{LO}_P$).
We fix the NLO operators to be the six local operators and the 
spin-orbit operator, as in Refs.~\cite{Gezerlis:2013ipa, Gezerlis:2014zia}: 
$\{\mathbbm{1},\sigma_{12},\tau_{12},\sigma_{12}\tau_{12},
({\bm \sigma}_1\cdot\vec{q})({\bm \sigma}_2\cdot\vec{q}),
({\bm \sigma}_1\cdot\vec{q})({\bm \sigma}_2\cdot\vec{q})\tau_{12}\}$ and
$({\bm \sigma}_1+{\bm \sigma}_2)\cdot(\vec{q}\times\vec{k})$.

In addition to the $S$-wave phase shifts, we fit 
the LECs to the $J=1$ mixing angle $\epsilon_1$ and the $^1P_1$ 
and $^3P_J$ partial waves. We give the NLO LECs for all interactions
in Table~\ref{tab:NLOLECs}.

We present the NLO phase shifts in the right panels in
Figs.~\ref{fig:SwavePS}--\ref{fig:DwavePS}. Regarding the $S$-wave
phase shifts, we find a good description of the PWA results. 
Again, all seven interactions produce the same phase shifts. In the
coupled channel, for the $J=1$ mixing angle and the $^3D_1$ phase
shift, we observe a dependence on the operator choice. The reason 
is that the LECs describing the NLO tensor-contact interactions are 
different for all operator pairs in order to reproduce the $P$-wave phase 
shifts. The different tensor interactions then lead to differences in the 
$^3D_1$ coupled channel, which affects the mixing angle. For the 
deuteron binding energy, we find an (almost) operator independent 
result of $E_d=2.113-2.134$ $(2.107-2.162)\mev$ for $R_0=1.0\fm$ 
$(1.2\fm)$, where the range is again due to the somewhat different tensor 
interactions.

Turning to the $P$ waves in Fig.~\ref{fig:PwavePS}, we observe two 
effects at NLO. First, the reproduction of the PWA values 
is much better for both the singlet and triplet partial waves. Second, 
we find that the effects of the violation of the Fierz rearrangement 
freedom are considerably reduced. At LO, we found a sizable spread 
of the description of the $P$-wave phase shifts already at energies around 
$20\mev$. At NLO, all interactions lead to similar phase shifts up to 
energies of $\sim100\mev$, but sizable regulator artifacts remain at higher
laboratory energies. The impact of the violation of the Fierz 
ambiguity is worst in the $^3P_2$ wave where the spin-orbit interaction 
is attractive and the tensor is weakest. Finally, for the higher-$l$ partial 
waves in Fig.~\ref{fig:DwavePS}, we find similar results for the phase 
shifts as at LO: These phase shifts are already well described by the 
OPE interaction at LO and improvements with the chiral order are counteracted
by the different tensor interactions. However, the violation of the Fierz 
rearrangement freedom has only a small impact on these partial waves.

We now discuss the effects on the many-body observables. For the 
\isotope[4]{He} ground-state energy we find that the spread for different 
LO operator pairs reduces considerably: from $8.7\mev$ at LO to 
$3.8\mev$ at NLO, ranging from $22.4\mev$ for the
$\mathbbm{1},\sigma_{12}$ interaction to $26.2\mev $ for the
$\tau_{12},\sigma_{12}\tau_{12}$ interaction. The spread is considerably 
smaller than the EKM uncertainties of at least $7\mev$, in contrast to 
the results at LO.

We show the neutron-matter energy at $n_0/2$ in the middle panel 
of Fig.~\ref{fig:NM4He}. In contrast to the results at LO, we now find
all interactions produce unbound neutron matter at roughly the same 
energy in the range of $9.1-9.7 \mev$. Thus, for neutron matter at 
$n_0/2$ the Fierz 
ambiguity is almost completely restored and the uncertainty from 
choosing different operator pairs is smaller than the EKM 
uncertainties. At higher density, $n_0$, the spread between different 
interactions remains larger at $3.0\mev$, ranging from $11.6\mev$ 
for the $\mathbbm{1}, \tau_{12}$ interaction up to $14.6\mev$ for the 
$\mathbbm{1}, \sigma_{12}$ interaction; see the right panel of 
Fig.~\ref{fig:NM4He}. This is to be expected, as the regulator artifacts 
increase with momentum or density (see also Refs.~\cite{Tews:2015ufa,
Dyhdalo:2016ygz}).

Considering only the interactions that gave reasonable results already 
at LO, the spread reduces to $1.5\mev$ and is, thus, only $1/4$
of the spread at LO ($5.8\mev$) and smaller than the EKM uncertainties 
of $\approx 2\mev$. The ordering of the NLO results closely follows the 
ordering of the description of the triplet $P$ waves, with the $\mathbbm{1}, 
\tau_{12}$ interaction being the biggest outlier.

In summary, we find that the violation of the Fierz ambiguity in the 
$NN$ sector is considerably reduced at NLO, and always covered 
by the EKM uncertainties, in contrast to the results at LO.
The violation of the Fierz ambiguity will remain of similar size at
N$^2$LO because no new contact terms are introduced, but we expect 
that it should be almost completely removed at N$^3$LO.
As a consequence of these findings, recently derived local $NN$
potentials at N$^2$LO can be used with confidence within their
uncertainties. 

Finally, we discuss implications for $3N$ interactions. We found
that the spread induced by different operator choices is sizable 
for the leading $NN$ interactions. This is also the case for the leading 
$3N$ interactions as shown in Refs.~\cite{Lynn:2015jua, Dyhdalo:2016ygz}. 
Our results demonstrate that the 
subleading contact interactions are necessary to reduce this spread. 
While in the $NN$ sector they appear already at NLO in chiral power
counting, the subleading $3N$ contact interactions only enter at 
N$^4$LO. The implementation of the N$^4$LO $3N$ forces is certainly 
challenging. Therefore, in order to tackle the 
violation of Fierz rearrangement freedom in the $3N$ sector, other ideas 
are necessary. These include the choice of the 
projected $V_E$ interaction of Ref.~\cite{Lynn:2015jua} or increasing 
the (momentum-space) cutoff for chiral interactions. However, we also 
note that we expect the EKM uncertainties to cover this effect for many 
(but not all) nuclear systems. In particular, for nuclear systems with 
densities (momenta) of the order of saturation density, we expect the 
EKM uncertainties to underestimate the effect coming from the violation 
of the Fierz ambiguity.

\section{Conclusion and Outlook} \label{seq:conclusion}

In summary, we investigated the violation of the Fierz rearrangement 
freedom that appears when using local regulators in the $NN$ sector 
at LO in chiral EFT.
We constructed interactions at LO and NLO in chiral EFT for all possible
pairs of LO operators where we fitted the LECs to the PWA results 
for the two $S$-wave phase-shift channels. We also constructed two 
complete LO interactions where we determined the additional two LECs 
either by projecting on partial waves with $S=T$ (attempting to restore
the Fierz rearrangement freedom), or by fitting to the $^1P_1$ and
$^3P_2$ phase shifts.

We studied phase shifts, the deuteron and \isotope[4]{He} binding 
energies, as well as the energy of neutron matter 
at different densities for all of these
interactions. Because we fitted the LECs to the two $S\neq T$
channels, we found that all interactions lead to a similar description
for these channels at LO. For the $S=T$ channels, i.e., the $^1P_1$ 
and $^3P_J$ partial waves, we found instead a strong dependence 
on the chosen operator structure for energies of the order of 
$20\mev$. At NLO, in addition to an improved description of the 
phase shifts in general, we found the effect of the violation of the
Fierz rearrangement freedom to be reduced considerably, having an effect 
only at energies larger than $100\mev$.

We then studied the \isotope[4]{He} ground-state energy, where the 
uncertainty due to the violation of the Fierz ambiguity reduced considerably 
when going from LO to NLO, from $8.7\mev$ to $3.8\mev$. In neutron 
matter, we found a sizable dependence on the chosen operator structure 
at LO. In particular, two interactions lead to bound neutron matter. For 
the five interactions that did not show a collapse, the spread from choosing 
different operator pairs reduced from $2.0\, (5.8)\mev$ to $0.2 \,(1.5)\mev$ 
when going from LO to NLO for $n=n_0/2 \,(n_0)$. At NLO, we found this 
spread to be much smaller than the EKM uncertainties for both densities. 
Furthermore, at saturation density, the operator dependence for the 
leading $NN$ interactions was found to be smaller than that for the leading 
$3N$ forces at N$^2$LO, with the latter being $\approx 4\mev$ for the 
three different operator choices in Ref.~\cite{Lynn:2015jua}.

We found that the violation of the Fierz ambiguity in the $NN$ sector is 
sizable at LO but restored to a large extent by including the subleading 
contact operators at NLO.
Any remaining violations of the Fierz ambiguity at higher orders in the
$NN$ sector are significantly smaller than the violation from the
leading $3N$ interaction at N$^2$LO.
The situation will further improve when contact interactions 
at N$^3$LO are included because additional 
correction terms will absorb the remaining regulator artifacts. Local chiral 
$NN$ interactions, thus, can be used with confidence even 
though additional regulator artifacts appear.

\begin{acknowledgments}
We thank S. Gandolfi for useful discussions. L.H. thanks the Institute for
Nuclear Theory in Seattle for its hospitality.
This work was supported by the ERC Grant No. 307986 STRONGINT, 
the National Science Foundation Grant No.~PHY-1430152 (JINA-CEE), 
the U.S.~DOE under Grant No.~DE-FG02-00ER41132, and the NUCLEI SciDAC
program.
Computational resources have been provided by the J\"ulich
Supercomputing Center, and the Lichtenberg high performance computer of
the TU Darmstadt.
We also used resources provided by NERSC, which
is supported by the U.S. DOE under Contract No.~DE-AC02-05CH11231.
\end{acknowledgments}

\bibliography{Paper_Fierz_v5}{}

\end{document}